\documentclass[prl,twocolumn]{revtex4-1}
\usepackage{graphicx}
\usepackage{amssymb}
\usepackage{bm}
\usepackage{amsmath,amsfonts,latexsym,braket}

\usepackage{hyperref}              
\usepackage{comment}
\usepackage{todonotes}

\newcommand*{\s}{\mathsf{S}}

\newcommand*{\cL}{{\mathcal L}}
\newcommand*{\cH}{{\mathcal H}}

\newcommand*{\cO}{{\mathcal O}}

\DeclareMathOperator{\tr}{tr}

\DeclareMathOperator{\E}{\mathbb{E}}

\begin{document}

\title{Noisy Spins and the Richardson--Gaudin Model}
\author{Daniel A. Rowlands}
\author{Austen Lamacraft}
\affiliation{TCM Group, Cavendish Laboratory, University of Cambridge, J. J. Thomson Ave., Cambridge CB3 0HE, UK}
\date{\today}
\email{dar55@cam.ac.uk}

\begin{abstract}
We study a system of spins (qubits) coupled to a common noisy environment, each precessing at its own frequency. The correlated noise experienced by the spins implies long-lived correlations that relax only due to the differing frequencies. We use a mapping to a non-Hermitian integrable Richardson--Gaudin model to find the exact spectrum of the quantum master equation in the high-temperature limit, and hence determine the decay rate. Our solution can be used to evaluate the effect of inhomogeneous splittings on a system of qubits coupled to a common bath.
\end{abstract}

\maketitle
The coherence of a quantum system is limited by the strength and nature of its coupling to the environment. Often, an environment consisting of many degrees of freedom can be treated as a source of noise that subjects the system to random disturbances \cite{Breuer:2002aa}. A central theme in quantum information science is the preparation and manipulation of quantum states in which such disturbance is minimal\cite{Zanardi:1997aa,Kempe:2001aa}.

The usual framework for the theoretical analysis of the open quantum systems described above is the quantum master equation (QME) for the system's density matrix $\rho$. Assuming Markovian dynamics, this may be written in Lindblad form \cite{Breuer:2002aa}
\begin{equation}\label{eq:QME}
  \dot\rho = -i\left[H,\rho\right] + \sum_\alpha \left[L_\alpha \rho L^\dagger_\alpha-\frac{1}{2}L^\dagger_\alpha L_\alpha \rho-\frac{1}{2}\rho L^\dagger_\alpha L_\alpha  \right],
\end{equation}
where $H$ is the system Hamiltonian, $L_\alpha$ are known as the Lindblad operators, and we set $\hbar=1$.

Solving the master equation exactly for a large system is, in general, impossible. However, as with pure unitary dynamics described by the Schr\"odinger equation, we may ask whether there are examples of exact solutions that are nontrivial, physically motivated, and valid for a system of arbitrary size. There is a long history of master equations of \emph{classical} stochastic processes being solved by methods developed for exactly solvable quantum models \cite{Golinelli:2006aa}. Surprisingly, very few examples of integrable QMEs -- allowing for a complete determination of the spectrum of decay modes -- may be found in the literature \cite{Prosen:2008aa,Prosen:2010aa,Medvedyeva:2016aa,Banchi:2017aa}.

In this Letter, we solve a model of $N$ spins described by \cite{Jeske:2013aa}
\begin{equation}\label{eq:model}
  \begin{split}
  H &= \sum_{j=1}^N \left[\Omega+\omega_j\right] s_j^z,\\
  L_z &= \sqrt{g_{0}}\sum_j s_j^z, \qquad L_\pm = \sqrt{g_\pm}\sum_j s_j^\pm
  \end{split}
\end{equation}
\begin{figure}
  \includegraphics[width = \columnwidth]{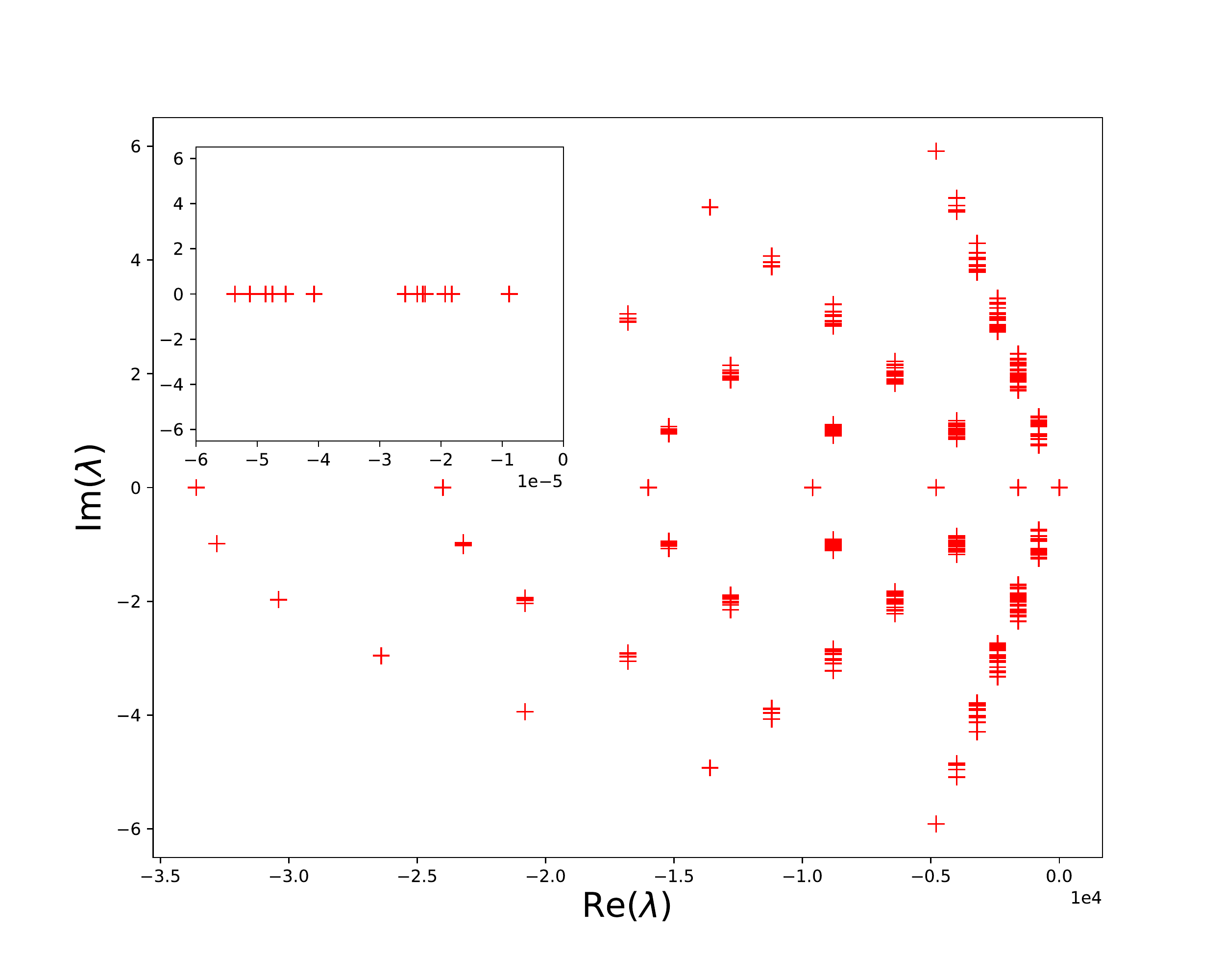}
  \caption{Spectrum of the Liouvillian (Eq.~\eqref{eq:RG-model}) for $n=6$ spin correlations for the case of $\Omega=1,g_+=800$, $g_0=0$ and $\omega_i \sim \text{Uni}(-0.2,0.2)$ obtained by exact diagonalization. The inset is a magnified view of a split multiplet of 15 states near zero. The spectrum is symmetric with respect to the real axis due to the PT symmetry of the Liouvillian.}
  \label{fig:omega-nonzero-spectrum}
\end{figure}
This model describes precession of the individual spins at frequencies $\Omega+\omega_i$, which could represent unequal level splittings in a system of qubits, for example. The $L_\alpha$ describe correlated coupling to the environment: $L_z$ accounts for pure dephasing, while $L_\pm$ describe the excitation and decay of the spins. The three couplings $g_0$, $g_\pm$ depend on the spectral density of the environment at frequencies $0$, $\pm \Omega$. Detailed balance for an environment at temperature $T$ implies $g_+/g_- = e^{-\Omega/k_\text{B}T}$. We solve the model Eq.~\eqref{eq:model} exactly in the high-temperature limit when $g_+=g_-$. This situation, describing incoherent driving, arises in many situations. As a representative sample, we cite superconducting qubits \cite{Devoret:2005aa}, photosynthetic light-harvesting complexes \cite{Haken:1973aa,Mohseni:2008aa,Trautmann:2017aa},
and ion traps \cite{Taylor:2017aa}. In a Rabi driven system, an infinite-temperature bath can arise as an effective description of a zero-temperature bath describing only spontaneous emission \cite{Hauss:2008aa}.

When $\omega_j=0$ the components of the density matrix describing isotropic spin correlations are stationary, corresponding to degenerate zero eigenvalues of the Liouvillian. The exact solution allows us to calculate the spectrum of $n$-spin correlations when $\omega_j\neq 0$ for arbitrary $n$, a result which can be obtained for only moderate $n$ by exact diagonalization (see Fig.~\ref{fig:omega-nonzero-spectrum}). When the $\omega_j$ are small, the decay rates have parametric form $\omega_j^2/g_+$, showing that increasing the noise reduces the decay rate, a manifestation of the quantum Zeno effect \cite{Beige:2000aa}. Although it is natural to interpret this in terms of second-order degenerate perturbation theory, it is not clear to us how to actually perform such a calculation. Indeed, the first step -- to resolve the degeneracy at $\omega_j=0$ into an appropriate eigenbasis -- is most effectively accomplished by the exact solution, with its many integrals of motion besides the Liouvillian.

Solving Eq.~\eqref{eq:model} is possible because of the correlated coupling to the environment. Models of this type may be traced back to Dicke's paper \cite{dicke1954coherence,Garraway:2011aa} on the spontaneous emission of atoms confined to a region smaller than the wavelength of the emitted light, and have appeared in many contexts since \cite{Jeske:2013aa}. Dicke identified superradiant and subradiant states of the atomic ensemble, corresponding to states of maximum and minimum total spin. For $\omega_j=0$, the QME may be written purely in terms of the total spin, and the solution was found long ago \cite{Belavin:1969aa,Agarwal:1970aa,Bonifacio:1971aa,Bonifacio:1971ab}. For $\omega_j\neq 0$ the total spin does not commute with the Hamiltonian. Our solution proceeds via a mapping to a non-Hermitian version of the Richardson--Gaudin model \cite{Dukelsky:2004aa}.


\emph{Density matrix and correlation functions}. For $s=1/2$ the density matrix for a single spin may be written
\begin{equation}
  \rho^{(1)} = \frac{1}{2}\openone + \mathbf{c}\cdot\boldsymbol{s},\qquad |\mathbf{c}|\leq 1,
\end{equation}
with $|\mathbf{c}|=1$ corresponding to pure states. More generally, a spin-s density matrix can be decomposed into a convex combination of spherical tensors $T^{(k)}_q$ ($k=0,1,\cdots,2s$ and $q=-k,-k+1,\cdots,k$) \cite{Fano:1957aa}.

For $N$ spins ($s=1/2$) we may write
\begin{equation}\label{eq:c_def}
  \rho^{(N)} = \frac{1}{2^N}\sum_{\{a_j\}} c_{a_1\cdots a_N} s_1^{a_1}\cdots s_N^{a_N},
\end{equation}
where $a_j=0,x,y,z$, with $s^0=\openone$. The coefficients $c_{a_1\cdots a_N}$ may be identified with the correlation functions of the spins
\begin{equation} \label{eq:correlator-components}
 c_{a_1\ldots a_N} = \tr\left[\rho^{(N)} s_1^{a_1}\cdots s_N^{a_N}\right].
\end{equation}
Note that $c_{0\ldots 0}=1$ is required by normalization of the density matrix. The reduced density matrix for any subsystem of spins is obtained by setting to zero the index for all spins in its complement.

\emph{Mapping to the Richardson--Gaudin model}. The equation of motion of $c_{a_1\cdots a_N}$ may be found by substituting Eq.~\eqref{eq:c_def} into the QME. First, we note that for $g_+=g_-$ we may write the Lindblad operators as
\begin{equation}
  L_{x,y} = \sqrt{g_{+}}\sum_j s_j^{x,y},\qquad L_z = \sqrt{g_0}\sum_j s_j^z.
\end{equation}
Considering now the effect of one of the $L_\alpha$ and invoking the cyclic invariance of the trace, we observe
\begin{equation} \label{eq:one-Lindblad-effect}
  \begin{split}
    &\sum_{j,k}\tr\left[s_k^{\alpha}\rho s_j^{\alpha}(\cdots)-\frac{1}{2}\{s_k^{\alpha}s_j^{\alpha},\rho\}(\cdots)\right] = \frac{1}{2} \sum_{j,k}\\ &\tr\left[\rho\Big(s_j^{\alpha}(\cdots)s_k^{\alpha} +s_k^{\alpha}(\cdots)s_j^{\alpha} \right.
     \left.- (\cdots)s_j^{\alpha}s_k^{\alpha} -s_j^{\alpha}s_k^{\alpha} (\cdots)\Big)\right].
  \end{split}
\end{equation}
We also note the following identity
\begin{equation}\label{eq:commutator}
  \begin{split}
    &s_j^{\alpha} s_j^{a_j} s_k^{a_k} s_k^{\alpha} +s_k^{\alpha}s_j^{a_j}s_k^{a_k}s_j^{\alpha}
   - s_j^{a_j}s_k^{a_k}s_j^{\alpha}s_k^{\alpha} -s_j^{\alpha}s_k^{\alpha} s_j^{a_j}s_k^{a_k} \\ &= -[s_j^{\alpha},s_j^{a_j}][s_k^{\alpha},s_k^{a_k}] \\
     &=\sum_{b,c} \varepsilon_{\alpha a_j b} \, \varepsilon_{\alpha a_k c} \, s_j^{b} s_k^{c}
     = (\mathsf{T}^{\alpha} \boldsymbol{s}_j)^{a_j}(\mathsf{T}^{\alpha} \boldsymbol{s}_k)^{a_k},
  \end{split}
\end{equation}
where $(\mathsf{T}^{\alpha})_{bc}=-\epsilon_{abc}$ are the generators of $\mathfrak{so}(3)$ in the adjoint representation. Since $\mathfrak{su}(2) \cong \mathfrak{so}(3)$, they can alternatively be thought of as generators of $\mathfrak{su}(2)$ in the adjoint representation.

If we switch to Hermitian Lie algebra generators, we can introduce spin-1 operators $\s_j^a = i\mathsf{T}_j^a$. After combining Eqs.~\eqref{eq:QME},\eqref{eq:one-Lindblad-effect}, and \eqref{eq:commutator}, we obtain the equation of motion for the correlator $\mathsf{C}$ (with tensor components defined by Eq.~\eqref{eq:correlator-components})
\begin{equation}
\partial_t \mathsf{C} = \cL \mathsf{C},
\label{eq:liouville}
\end{equation}
where the Liouvillian superoperator $\cL$ takes the form of the non-Hermitian spin-1 Richardson-Gaudin model
\begin{equation} \label{eq:RG-model}
\cL = i\sum_{j=1}^n\left[\Omega+\omega_j\right]\s^z_j - g_+\sum_{j,k=1}^n \left(\s^x_j \s^x_k+\s^y_j \s^y_k\right) - g_{0}\sum_{j,k=1}^n \s^z_j \s^z_k.
\end{equation}
Here $n$ is the number of nonzero indices of $\mathsf{C}$, which describe the reduced density matrix of the corresponding spins. The same model, involving a system of spins with $\mathsf{S}_j=1,\ldots, 2s$, would arise for spin-s physical degrees of freedom.

\emph{Equivalence to stochastic evolution}. We can obtain the same result in a more robust and transparent fashion by regarding the high-temperature limit ($g_+ = g_-$) as a problem of stochastic evolution due to classical noise \cite{Gardiner:1985aa,Barchielli:1986aa,Ghirardi:1990aa,Adler:2000aa,Bauer:2013aa,Bauer:2017aa,Chenu:2017aa}.

Consider $N$ spins precessing in a common stochastic field, so that their evolution is governed by the Hamiltonian $H_\eta=\sum_{j=1}^N h_j(t)$, where
\begin{equation}
  h_j(t) = \eta_x(t) s^x_j + \eta_y(t) s^y_j+\left[\Omega+\omega_j + \eta_z(t)\right]s^z_j,
\end{equation}
and $\eta_j(t)$ describe Gaussian white noises with covariances $\E[\eta_z(t)\eta_z(t')] = g_0 \delta(t-t')$ and $\E[\eta_x(t)\eta_x(t')]=\E[\eta_y(t)\eta_y(t')]=g_+\delta(t-t')$.
The corresponding infinitesimal stochastic unitary evolution $U(t+dt,t) = e^{-i\mathrm{d}H_t}$ is generated by
\begin{equation}
\mathrm{d}H_t = \sum_j (\Omega + \omega_j)s_j^z \; \mathrm{d}t + \sum_{j,\alpha} \sqrt{g_\alpha} \, s_j^\alpha \; \mathrm{d}\eta^\alpha_t,
\end{equation}
from which it follows by It\^o's lemma that the density matrix $\varrho_t$ satisfies the It\^o stochastic differential equation
\begin{equation}
\mathrm{d}\varrho_t = -\left(i[H,\varrho_t] + \frac{1}{2}\sum_\alpha [L_\alpha,[L_\alpha,\varrho_t]]\right)\mathrm{d}t - i\sum_\alpha [L_\alpha,\varrho_t] \mathrm{d}\eta^\alpha_t.
\end{equation}
%
%
%
%
After averaging, $\rho = \E_\eta[ \varrho]$ can be seen to satisfy the QME described by Eq. \eqref{eq:model}. However, we could alternatively consider the evolution of the correlation tensor $\mathsf{C}$, which for \emph{non-stochastic} $\eta_j$ would be given by Eq. \eqref{eq:liouville} with
\begin{equation}
i\cL_\text{ns} = \sum_{j=1}^n \eta_x(t) \s^x_j + \eta_y(t)\s^y_j+\left[\Omega+\omega_j + \eta_z(t)\right]\s^z_j.
\end{equation}
Stochastic $\eta_j$ therefore gives rise to It\^o terms describing the spin-spin interaction in Eq. \eqref{eq:RG-model}.

\emph{Exact solution}.
As a prelude to the exact solution of Eq.~\eqref{eq:RG-model}, we first consider the much simpler case of $\omega_j=0$ (and $g_0=0$), such that the model reduces to
\begin{equation}
\cL = i\Omega \s^z_{\text{tot}} - g \left[\mathbf{\mathsf{S}}^2_{\text{tot}} - (\s^z_{\text{tot}})^2\right],
\end{equation}
from which the spectrum can be obtained immediately. It consists of degenerate multiplets for given values of $(\mathsf{S}_{\text{tot}},\mathsf{S}^z_{\text{tot}})$, with the multiplets of fixed $\mathsf{S}_{\text{tot}}$ lying on parabolas. In particular, states with $\mathsf{S}_{\text{tot}}=0$ have exactly zero eigenvalue. For these states, the tensor $c_{a_1\ldots a_N}$ is \emph{isotropic}. The simplest example is provided by $N=2$, where the most general rotationally invariant density matrix (two-qubit Werner state) is
\begin{equation}
  \rho^{(2)}_c = \frac{1}{4}\openone + c_{\bullet}\,\textbf{s}_1\cdot \textbf{s}_2,\qquad -1\leq c_\bullet\leq 1/3,
\end{equation}
corresponding to $c_{00}=1$, and $c_{a_1a_2} = 4c_\bullet\delta_{a_1,a_2}$ for $a_{1,2}=x,y,z$. Note that $c_\bullet=-1$ corresponds to a pure singlet state, but for larger $N$ one cannot express the isotropic tensors only in terms of singlet states. By virtue of the Choi isomorphism, the density matrix can be regarded as an element of the tensor product space $\cH\otimes \cH$, where $\cH=(\mathbb{C}^2)^{\otimes N}$ is the Hilbert space of $N$ spins. Thus the isotropic tensors with up to $N$ indices are the $\mathsf{S}_\text{tot}=0$ states formed from $2N$ spin-1/2s, which number $\frac{1}{N+1}\binom{2N}{N}$ (the Catalan numbers, $C_N$). The number of isotropic tensors of fixed rank $n$ is the number of $\mathsf{S}_\text{tot}=0$ states that can be formed from $n$ spin-1s. These are the Riordan numbers $R_n= \sum_{m=0}^n (-1)^{n-m} \binom{n}{m} C_m$  \cite{Bernhart:1999aa,Sloane:2017aa,Andrews:1977aa}. 

Turning to nonzero $\omega_i$, the multiplets can be seen to split as shown in Fig.~\ref{fig:omega-nonzero-spectrum}. To find the decay rate, one must identify the state whose eigenvalue has the least negative real part (which we shall term the dominant eigenvalue). Therefore, for small $\omega_i$ $(|\omega_i| \ll |\Omega|)$ at least, the dominant eigenvalue will lie within the $\s_{\text{tot}}=0$ (i.e. singlet) subspace. The splitting of the singlet multiplet in the real direction can be thought of as a second-order perturbative correction of the form $\omega_i^2/g_+$. However, for this problem we are in fact afforded a more facile route via the exact solution, to which we now turn.

The exact eigenstates of Eq.~\eqref{eq:RG-model} take the Bethe form \cite{Links:2003}
\begin{equation} \label{eq:Bethe-state}
\lvert\mu_1\cdots \mu_m\rangle = \prod_{k=1}^m \left(\sum_{j=1}^n \frac{\s^+_j}{\mu_k - \frac{1}{2}i\omega_j}\right) \ket{\chi^-},
\end{equation}
where $\s_{\text{tot}}^z= m-n$, the pseudovacuum $\ket{\chi^-}$ is the lowest weight state $\ket{-1}^{\otimes n}$, and the Bethe roots $\{\mu_i\}$ satisfy the Bethe ansatz equations
\begin{equation} \label{eq:Bethe-equations}
  \frac{1}{g_+} + \sum_{k=1}^n \frac{1}{\mu_j- \frac{1}{2}i\omega_k} - \sum_{k\neq j}^m\frac{1}{\mu_j-\mu_k} = 0.
\end{equation}
The eigenvalue $\lambda(\vec{\mu})$ of a Bethe state is given by
\begin{equation}
\lambda(\vec{\mu}) = 2 \sum_{j=1}^m \mu_j - i \sum_{j=1}^n \omega_j,
\end{equation}
where, since $\s_{\text{tot}}^z$ is conserved, we continue to set $g_0=0$ without loss of generality.

Equations \eqref{eq:Bethe-equations} can be interpreted in terms of two-dimensional classical electrostatics \cite{Dukelsky:2004aa}: if the $\omega_i$ and $\mu_i$ correspond to the positions of fixed and free point charges respectively, and $1/g_+$ represents a uniform electric field, then Eq.~\eqref{eq:Bethe-equations} describes the equilibrium condition. The equilibrium configurations describe saddle points of the energy (Earnshaw's theorem), and so finding all solutions for large $n$ is a difficult task.

\begin{figure}
  \includegraphics[width = \columnwidth]{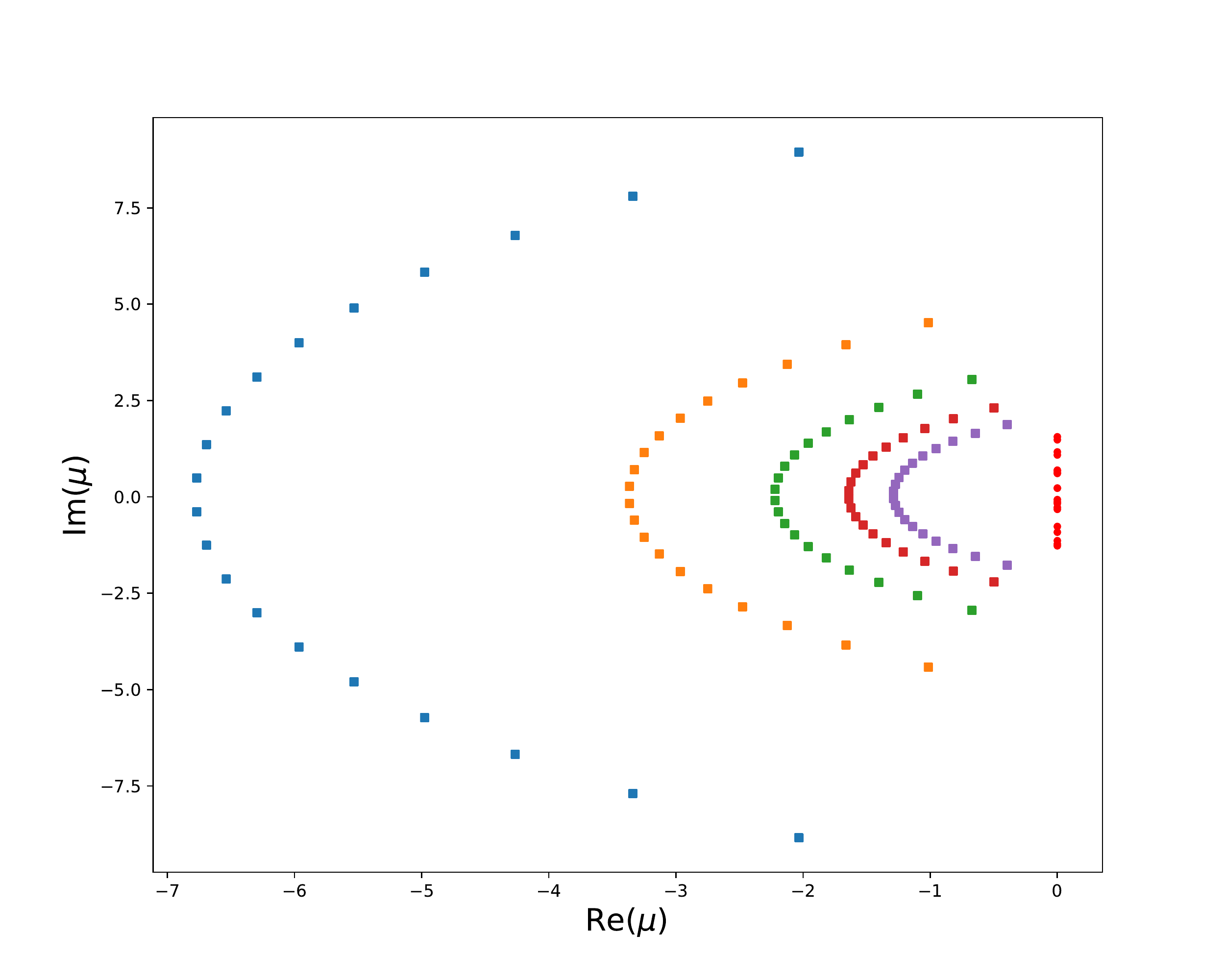}
  \caption{Bethe root distribution corresponding to the $\s_{\text{tot}}^z=0$ eigenstate descended from the maximal $\s_{\text{tot}}$ state of the $\omega_i=0$ model ($n=20$). The curves of different colour correspond to different values of $1/g_+$ (increasing from left to right), and the $\omega_i$ are shown as red circles along the imaginary axis. One can see that $\s_{\text{tot}}$ is maximum by noting that all $\mu_i$ go to infinity as $1/g_+$ vanishes, and so from Eq.~\eqref{eq:Bethe-state} the state is derived from $\ket{\chi^-}$ simply by raising $\s_{\text{tot}}^z$ to zero.}
  \label{fig:S_max}
\end{figure}

Naive numerical root finding on the Bethe equations for random $\omega_j$ configurations tends to yield solutions in which the Bethe roots condense onto curves as shown in Fig.~\ref{fig:S_max}. These are the descendants of the states of maximum $\s_{\text{tot}}$ (when $\omega_i=0$), which though of interest in the context of superradiance do not directly concern us here. We note in passing that the analogue of superradiance that appears here is that the eigenvalues of these states (for fixed $g_+$) scale quadratically with $n$, and so the correlations for states of $\s_{\text{tot}} \sim n$ decay at a rate that is $\cO(n^2)$. This is to be contrasted with the $\cO(n)$ decay rate of the singlet correlations, which we shall discuss next.

\begin{figure}
  \includegraphics[width = \columnwidth]{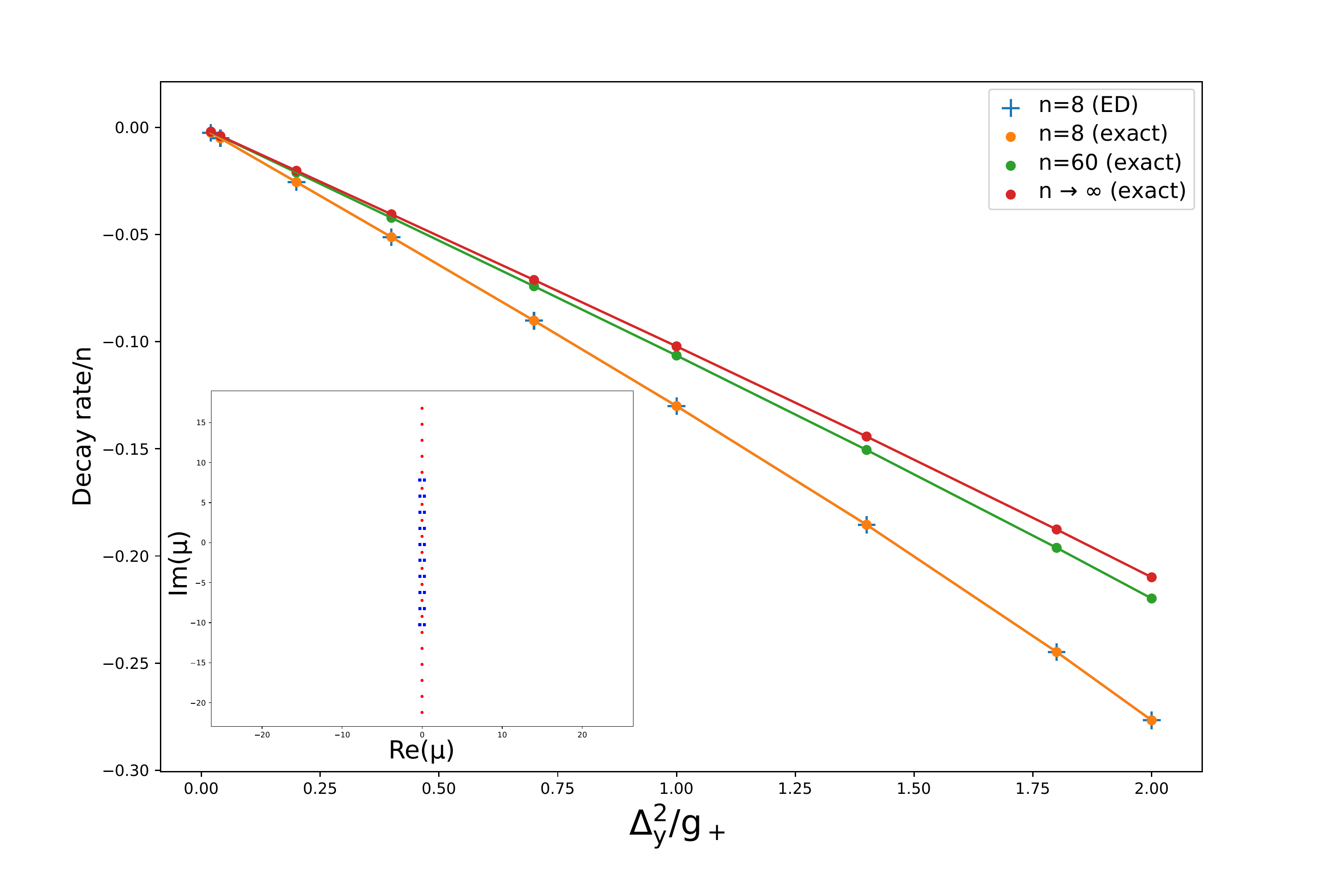}
  \caption{Comparison of the decay rate (dominant Liouvillian eigenvalue) of $n$-spin correlations for $\Omega=-(n+3)$ and $\omega_j= j\Delta_y$ (where $\Delta_y=2$) evaluated by (i) exact diagonalization for $n=8$, (ii) exact solution at $n=8$ and $60$ and (iii) exact solution for $n\to \infty$. The inset shows the string state formed by the Bethe roots (blue squares), found by numerical solution of Eq.~\eqref{eq:Bethe-equations} for $n=20$ and $g_+ \to \infty$; as in Fig.~\ref{fig:S_max}, the $\omega_i$ are represented by red circles. The effect of finite $g_+$ is to push the Bethe roots in the negative real direction.}
  \label{fig:compare}
\end{figure}

We were able to find the Bethe roots for the dominant state in the case of uniformly spaced $\omega_i$: they form the string state shown in the inset in Fig.~\ref{fig:compare}. In the $n \to \infty$ limit, it is possible to evaluate the infinite summations in Eq.~\eqref{eq:Bethe-equations} exactly. If the spacing of the fixed charges is $i\Delta_y$ and the free charges on either side of the imaginary axis have real parts $\Delta_+$ and $-\Delta_-$, we are left with
\begin{equation}
  \begin{split}
  \frac{2\pi}{\Delta_y}\tanh\left(\frac{2\pi \Delta_+}{\Delta_y}\right) = \frac{\pi}{\Delta_y}\coth\left(\frac{\pi (\Delta_++\Delta_-)}{\Delta_y}\right)-\frac{1}{g_+},\\
  \frac{2\pi}{\Delta_y}\tanh\left(\frac{2\pi \Delta_-}{\Delta_y}\right) = \frac{\pi}{\Delta_y}\coth\left(\frac{\pi (\Delta_++\Delta_-)}{\Delta_y}\right)+\frac{1}{g_+}.
  \end{split}
\end{equation}
Solving these two equations numerically for $\Delta_\pm$ enables us to find the Liouvillian eigenvalue of the string state.

In Fig.~\ref{fig:compare}, we show convergence of the finite $n$ solution of the Bethe equations to this large $n$ result and also verify that, for small $n$, the string solution coincides with the dominant eigenvalue found by exact diagonalization. The observed linear dependence of the dominant eigenvalue on $1/g_+$ is consistent with the aforementioned $\omega_i^2/g_+$ splitting predicted by perturbation theory.

\begin{figure}
  \includegraphics[width = \columnwidth]{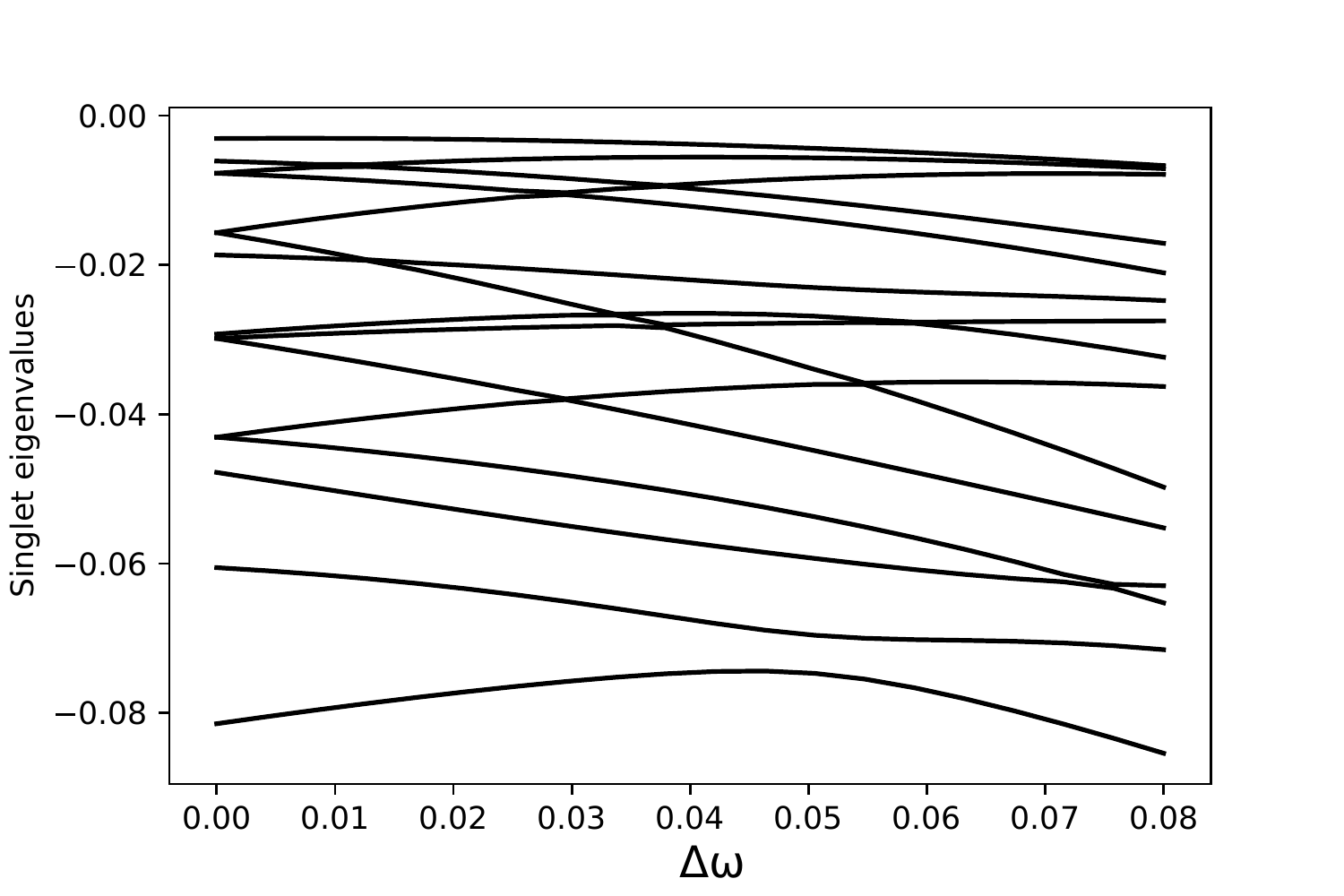}
  \caption{Typical motion of the singlet eigenvalues as the $\omega_i$ are smoothly translated (parameterised by $\Delta \omega$), revealing the presence of level crossings.}
  \label{fig:poisson}
\end{figure}

A further interesting consequence of the integrability of our model is the absence of level repulsion as the spectrum varies with varying $\omega_i$ (see Fig.~\ref{fig:poisson}), leading to Poissonian level statistics. We conjecture that choosing $\omega_i$ to be independent and identically distributed will therefore lead to the relaxation rate (magnitude of the real part of the dominant eigenvalue)  $\lambda_0$ $(\lambda_0 \geq 0)$ having the Weibull distribution $\frac{\alpha}{\beta} \left(\frac{\lambda_0}{\beta}\right)^{\alpha-1} e^{-(\lambda_0/\beta)^\alpha}$ for some $\alpha$ and $\beta$ \cite{Gumbel:2012aa}.

\emph{Conclusions and outlook}. We have computed the exact relaxation rate of correlations in a model of spins precessing at different frequencies and coupled to a common noise source by exploiting a mapping to an exactly solvable model in the high-temperature limit. Our solution can be used to evaluate the effect of inhomogeneous splittings on a system of qubits coupled to a common bath.

The derivation of the spin-spin interaction in \eqref{eq:RG-model} may be generalized to the case of noise with arbitrary correlations between different spins $j$ and $k$, leading to a coupling $g_{ij}$ that could define an arbitrary quadratic spin-spin interaction. In general, the dominant eigenvalue of such an interaction will be nonzero and negative -- a spin model will have a finite positive ground state energy -- whereas for the infinite-range coupling we have considered, a nonzero dominant eigenvalue arises because of the $\omega_i$. Nevertheless, it would be interesting to explore other possibilities, e.g., integrable 1D spin chains.

What happens at finite temperature when $g_+ \not= g_-$ -- a situation describing relaxation as well as classical noise? The Lindblad operators vanish on any state $\ket{\Psi}$ satisfying $\sum_j\mathbf{s}_{j}\ket{\Psi}=0$, and for $\omega_i=0$ these form a decoherence free subspace for $N$ even of dimension $C_{N/2}$, for any $g_\pm$  \cite{Zanardi:1997aa,Kempe:2001aa}. Density matrices formed from these states are a subset of the isotropic density matrices considered earlier. As in that case, $\omega_j\neq 0$ will cause decoherence of this subspace. Unfortunately, we have no reason to believe that the model remains integrable in the more general case, so finding an analytical description of the relaxation of $n$-spin correlations at finite temperature remains an open problem.

\noindent\emph{Acknowledgments}. DAR and AL gratefully acknowledge the EPSRC for financial support, under Grants No. EP/M506485/1 and No. EP/P034616/1 respectively.

\bibliographystyle{apsrev}
\bibliography{bibliography.bib}

\begin{thebibliography}{36}
\expandafter\ifx\csname natexlab\endcsname\relax\def\natexlab#1{#1}\fi
\expandafter\ifx\csname bibnamefont\endcsname\relax
  \def\bibnamefont#1{#1}\fi
\expandafter\ifx\csname bibfnamefont\endcsname\relax
  \def\bibfnamefont#1{#1}\fi
\expandafter\ifx\csname citenamefont\endcsname\relax
  \def\citenamefont#1{#1}\fi
\expandafter\ifx\csname url\endcsname\relax
  \def\url#1{\texttt{#1}}\fi
\expandafter\ifx\csname urlprefix\endcsname\relax\def\urlprefix{URL }\fi
\providecommand{\bibinfo}[2]{#2}
\providecommand{\eprint}[2][]{\url{#2}}

\bibitem[{\citenamefont{Breuer and Petruccione}(2002)}]{Breuer:2002aa}
\bibinfo{author}{\bibfnamefont{H.-P.} \bibnamefont{Breuer}} \bibnamefont{and}
  \bibinfo{author}{\bibfnamefont{F.}~\bibnamefont{Petruccione}},
  \emph{\bibinfo{title}{The theory of open quantum systems}}
  (\bibinfo{publisher}{Oxford University Press on Demand},
  \bibinfo{year}{2002}).

\bibitem[{\citenamefont{Zanardi and Rasetti}(1997)}]{Zanardi:1997aa}
\bibinfo{author}{\bibfnamefont{P.}~\bibnamefont{Zanardi}} \bibnamefont{and}
  \bibinfo{author}{\bibfnamefont{M.}~\bibnamefont{Rasetti}},
  \bibinfo{journal}{Physical Review Letters} \textbf{\bibinfo{volume}{79}},
  \bibinfo{pages}{3306} (\bibinfo{year}{1997}).

\bibitem[{\citenamefont{Kempe et~al.}(2001)\citenamefont{Kempe, Bacon, Lidar,
  and Whaley}}]{Kempe:2001aa}
\bibinfo{author}{\bibfnamefont{J.}~\bibnamefont{Kempe}},
  \bibinfo{author}{\bibfnamefont{D.}~\bibnamefont{Bacon}},
  \bibinfo{author}{\bibfnamefont{D.~A.} \bibnamefont{Lidar}}, \bibnamefont{and}
  \bibinfo{author}{\bibfnamefont{K.~B.} \bibnamefont{Whaley}},
  \bibinfo{journal}{Physical Review A} \textbf{\bibinfo{volume}{63}},
  \bibinfo{pages}{042307} (\bibinfo{year}{2001}).

\bibitem[{\citenamefont{Golinelli and Mallick}(2006)}]{Golinelli:2006aa}
\bibinfo{author}{\bibfnamefont{O.}~\bibnamefont{Golinelli}} \bibnamefont{and}
  \bibinfo{author}{\bibfnamefont{K.}~\bibnamefont{Mallick}},
  \bibinfo{journal}{Journal of Physics A: Mathematical and General}
  \textbf{\bibinfo{volume}{39}}, \bibinfo{pages}{12679} (\bibinfo{year}{2006}).

\bibitem[{\citenamefont{Prosen}(2008)}]{Prosen:2008aa}
\bibinfo{author}{\bibfnamefont{T.}~\bibnamefont{Prosen}}, \bibinfo{journal}{New
  Journal of Physics} \textbf{\bibinfo{volume}{10}}, \bibinfo{pages}{043026}
  (\bibinfo{year}{2008}).

\bibitem[{\citenamefont{Prosen and Seligman}(2010)}]{Prosen:2010aa}
\bibinfo{author}{\bibfnamefont{T.}~\bibnamefont{Prosen}} \bibnamefont{and}
  \bibinfo{author}{\bibfnamefont{T.~H.} \bibnamefont{Seligman}},
  \bibinfo{journal}{Journal of Physics A: Mathematical and Theoretical}
  \textbf{\bibinfo{volume}{43}}, \bibinfo{pages}{392004}
  (\bibinfo{year}{2010}).

\bibitem[{\citenamefont{Medvedyeva et~al.}(2016)\citenamefont{Medvedyeva,
  Essler, and Prosen}}]{Medvedyeva:2016aa}
\bibinfo{author}{\bibfnamefont{M.~V.} \bibnamefont{Medvedyeva}},
  \bibinfo{author}{\bibfnamefont{F.~H.} \bibnamefont{Essler}},
  \bibnamefont{and} \bibinfo{author}{\bibfnamefont{T.}~\bibnamefont{Prosen}},
  \bibinfo{journal}{Physical Review Letters} \textbf{\bibinfo{volume}{117}},
  \bibinfo{pages}{137202} (\bibinfo{year}{2016}).

\bibitem[{\citenamefont{Banchi et~al.}(2017)\citenamefont{Banchi, Burgarth, and
  Kastoryano}}]{Banchi:2017aa}
\bibinfo{author}{\bibfnamefont{L.}~\bibnamefont{Banchi}},
  \bibinfo{author}{\bibfnamefont{D.}~\bibnamefont{Burgarth}}, \bibnamefont{and}
  \bibinfo{author}{\bibfnamefont{M.~J.} \bibnamefont{Kastoryano}},
  \bibinfo{journal}{Phys. Rev. X} \textbf{\bibinfo{volume}{7}},
  \bibinfo{pages}{041015} (\bibinfo{year}{2017}),
  \urlprefix\url{https://link.aps.org/doi/10.1103/PhysRevX.7.041015}.

\bibitem[{\citenamefont{Jeske and Cole}(2013)}]{Jeske:2013aa}
\bibinfo{author}{\bibfnamefont{J.}~\bibnamefont{Jeske}} \bibnamefont{and}
  \bibinfo{author}{\bibfnamefont{J.~H.} \bibnamefont{Cole}},
  \bibinfo{journal}{Phys. Rev. A} \textbf{\bibinfo{volume}{87}},
  \bibinfo{pages}{052138} (\bibinfo{year}{2013}),
  \urlprefix\url{https://link.aps.org/doi/10.1103/PhysRevA.87.052138}.

\bibitem[{\citenamefont{Devoret and Martinis}(2005)}]{Devoret:2005aa}
\bibinfo{author}{\bibfnamefont{M.~H.} \bibnamefont{Devoret}} \bibnamefont{and}
  \bibinfo{author}{\bibfnamefont{J.~M.} \bibnamefont{Martinis}}, in
  \emph{\bibinfo{booktitle}{Experimental aspects of quantum computing}}
  (\bibinfo{publisher}{Springer}, \bibinfo{year}{2005}), pp.
  \bibinfo{pages}{163--203}.

\bibitem[{\citenamefont{Haken and Strobl}(1973)}]{Haken:1973aa}
\bibinfo{author}{\bibfnamefont{H.}~\bibnamefont{Haken}} \bibnamefont{and}
  \bibinfo{author}{\bibfnamefont{G.}~\bibnamefont{Strobl}},
  \bibinfo{journal}{Zeitschrift f{\"u}r Physik A Hadrons and Nuclei}
  \textbf{\bibinfo{volume}{262}}, \bibinfo{pages}{135} (\bibinfo{year}{1973}).

\bibitem[{\citenamefont{Mohseni et~al.}(2008)\citenamefont{Mohseni, Rebentrost,
  Lloyd, and Aspuru-Guzik}}]{Mohseni:2008aa}
\bibinfo{author}{\bibfnamefont{M.}~\bibnamefont{Mohseni}},
  \bibinfo{author}{\bibfnamefont{P.}~\bibnamefont{Rebentrost}},
  \bibinfo{author}{\bibfnamefont{S.}~\bibnamefont{Lloyd}}, \bibnamefont{and}
  \bibinfo{author}{\bibfnamefont{A.}~\bibnamefont{Aspuru-Guzik}},
  \bibinfo{journal}{The Journal of Chemical Physics}
  \textbf{\bibinfo{volume}{129}}, \bibinfo{pages}{174106}
  (\bibinfo{year}{2008}).

\bibitem[{\citenamefont{Trautmann and Hauke}(2017)}]{Trautmann:2017aa}
\bibinfo{author}{\bibfnamefont{N.}~\bibnamefont{Trautmann}} \bibnamefont{and}
  \bibinfo{author}{\bibfnamefont{P.}~\bibnamefont{Hauke}},
  \bibinfo{journal}{Physical Review A} \textbf{\bibinfo{volume}{97}},
  \bibinfo{pages}{0236} (\bibinfo{year}{2017}).

\bibitem[{\citenamefont{Taylor et~al.}(2017)\citenamefont{Taylor, Bentley,
  Pedernales, Lamata, Solano, Carvalho, and Hope}}]{Taylor:2017aa}
\bibinfo{author}{\bibfnamefont{R.~L.} \bibnamefont{Taylor}},
  \bibinfo{author}{\bibfnamefont{C.~D.} \bibnamefont{Bentley}},
  \bibinfo{author}{\bibfnamefont{J.~S.} \bibnamefont{Pedernales}},
  \bibinfo{author}{\bibfnamefont{L.}~\bibnamefont{Lamata}},
  \bibinfo{author}{\bibfnamefont{E.}~\bibnamefont{Solano}},
  \bibinfo{author}{\bibfnamefont{A.~R.} \bibnamefont{Carvalho}},
  \bibnamefont{and} \bibinfo{author}{\bibfnamefont{J.~J.} \bibnamefont{Hope}},
  \bibinfo{journal}{Scientific Reports} \textbf{\bibinfo{volume}{7}},
  \bibinfo{pages}{46197} (\bibinfo{year}{2017}).

\bibitem[{\citenamefont{Hauss et~al.}(2008)\citenamefont{Hauss, Fedorov,
  Hutter, Shnirman, and Sch{\"o}n}}]{Hauss:2008aa}
\bibinfo{author}{\bibfnamefont{J.}~\bibnamefont{Hauss}},
  \bibinfo{author}{\bibfnamefont{A.}~\bibnamefont{Fedorov}},
  \bibinfo{author}{\bibfnamefont{C.}~\bibnamefont{Hutter}},
  \bibinfo{author}{\bibfnamefont{A.}~\bibnamefont{Shnirman}}, \bibnamefont{and}
  \bibinfo{author}{\bibfnamefont{G.}~\bibnamefont{Sch{\"o}n}},
  \bibinfo{journal}{Physical Review Letters} \textbf{\bibinfo{volume}{100}},
  \bibinfo{pages}{037003} (\bibinfo{year}{2008}).

\bibitem[{\citenamefont{Beige et~al.}(2000)\citenamefont{Beige, Braun,
  Tregenna, and Knight}}]{Beige:2000aa}
\bibinfo{author}{\bibfnamefont{A.}~\bibnamefont{Beige}},
  \bibinfo{author}{\bibfnamefont{D.}~\bibnamefont{Braun}},
  \bibinfo{author}{\bibfnamefont{B.}~\bibnamefont{Tregenna}}, \bibnamefont{and}
  \bibinfo{author}{\bibfnamefont{P.~L.} \bibnamefont{Knight}},
  \bibinfo{journal}{Physical Review Letters} \textbf{\bibinfo{volume}{85}},
  \bibinfo{pages}{1762} (\bibinfo{year}{2000}).

\bibitem[{\citenamefont{Dicke}(1954)}]{dicke1954coherence}
\bibinfo{author}{\bibfnamefont{R.~H.} \bibnamefont{Dicke}},
  \bibinfo{journal}{Physical Review} \textbf{\bibinfo{volume}{93}},
  \bibinfo{pages}{99} (\bibinfo{year}{1954}).

\bibitem[{\citenamefont{Garraway}(2011)}]{Garraway:2011aa}
\bibinfo{author}{\bibfnamefont{B.~M.} \bibnamefont{Garraway}},
  \bibinfo{journal}{Philosophical Transactions of the Royal Society of London
  A: Mathematical, Physical and Engineering Sciences}
  \textbf{\bibinfo{volume}{369}}, \bibinfo{pages}{1137} (\bibinfo{year}{2011}).

\bibitem[{\citenamefont{Belavin et~al.}(1969)\citenamefont{Belavin, Zeldovich,
  Perelomov, and Popov}}]{Belavin:1969aa}
\bibinfo{author}{\bibfnamefont{A.}~\bibnamefont{Belavin}},
  \bibinfo{author}{\bibfnamefont{B.~Y.} \bibnamefont{Zeldovich}},
  \bibinfo{author}{\bibfnamefont{A.}~\bibnamefont{Perelomov}},
  \bibnamefont{and} \bibinfo{author}{\bibfnamefont{V.}~\bibnamefont{Popov}},
  \bibinfo{journal}{Sov. Phys. JETP} \textbf{\bibinfo{volume}{56}},
  \bibinfo{pages}{264} (\bibinfo{year}{1969}).

\bibitem[{\citenamefont{Agarwal}(1970)}]{Agarwal:1970aa}
\bibinfo{author}{\bibfnamefont{G.}~\bibnamefont{Agarwal}},
  \bibinfo{journal}{Physical Review A} \textbf{\bibinfo{volume}{2}},
  \bibinfo{pages}{2038} (\bibinfo{year}{1970}).

\bibitem[{\citenamefont{Bonifacio
  et~al.}(1971{\natexlab{a}})\citenamefont{Bonifacio, Schwendimann, and
  Haake}}]{Bonifacio:1971aa}
\bibinfo{author}{\bibfnamefont{R.}~\bibnamefont{Bonifacio}},
  \bibinfo{author}{\bibfnamefont{P.}~\bibnamefont{Schwendimann}},
  \bibnamefont{and} \bibinfo{author}{\bibfnamefont{F.}~\bibnamefont{Haake}},
  \bibinfo{journal}{Physical Review A} \textbf{\bibinfo{volume}{4}},
  \bibinfo{pages}{302} (\bibinfo{year}{1971}{\natexlab{a}}).

\bibitem[{\citenamefont{Bonifacio
  et~al.}(1971{\natexlab{b}})\citenamefont{Bonifacio, Schwendimann, and
  Haake}}]{Bonifacio:1971ab}
\bibinfo{author}{\bibfnamefont{R.}~\bibnamefont{Bonifacio}},
  \bibinfo{author}{\bibfnamefont{P.}~\bibnamefont{Schwendimann}},
  \bibnamefont{and} \bibinfo{author}{\bibfnamefont{F.}~\bibnamefont{Haake}},
  \bibinfo{journal}{Physical Review A} \textbf{\bibinfo{volume}{4}},
  \bibinfo{pages}{854} (\bibinfo{year}{1971}{\natexlab{b}}).

\bibitem[{\citenamefont{Dukelsky et~al.}(2004)\citenamefont{Dukelsky, Pittel,
  and Sierra}}]{Dukelsky:2004aa}
\bibinfo{author}{\bibfnamefont{J.}~\bibnamefont{Dukelsky}},
  \bibinfo{author}{\bibfnamefont{S.}~\bibnamefont{Pittel}}, \bibnamefont{and}
  \bibinfo{author}{\bibfnamefont{G.}~\bibnamefont{Sierra}},
  \bibinfo{journal}{Reviews of modern physics} \textbf{\bibinfo{volume}{76}},
  \bibinfo{pages}{643} (\bibinfo{year}{2004}).

\bibitem[{\citenamefont{Fano}(1957)}]{Fano:1957aa}
\bibinfo{author}{\bibfnamefont{U.}~\bibnamefont{Fano}},
  \bibinfo{journal}{Reviews of Modern Physics} \textbf{\bibinfo{volume}{29}},
  \bibinfo{pages}{74} (\bibinfo{year}{1957}).

\bibitem[{\citenamefont{Gardiner and Collett}(1985)}]{Gardiner:1985aa}
\bibinfo{author}{\bibfnamefont{C.}~\bibnamefont{Gardiner}} \bibnamefont{and}
  \bibinfo{author}{\bibfnamefont{M.}~\bibnamefont{Collett}},
  \bibinfo{journal}{Physical Review A} \textbf{\bibinfo{volume}{31}},
  \bibinfo{pages}{3761} (\bibinfo{year}{1985}).

\bibitem[{\citenamefont{Barchielli}(1986)}]{Barchielli:1986aa}
\bibinfo{author}{\bibfnamefont{A.}~\bibnamefont{Barchielli}},
  \bibinfo{journal}{Physical Review A} \textbf{\bibinfo{volume}{34}},
  \bibinfo{pages}{1642} (\bibinfo{year}{1986}).

\bibitem[{\citenamefont{Ghirardi et~al.}(1990)\citenamefont{Ghirardi, Pearle,
  and Rimini}}]{Ghirardi:1990aa}
\bibinfo{author}{\bibfnamefont{G.~C.} \bibnamefont{Ghirardi}},
  \bibinfo{author}{\bibfnamefont{P.}~\bibnamefont{Pearle}}, \bibnamefont{and}
  \bibinfo{author}{\bibfnamefont{A.}~\bibnamefont{Rimini}},
  \bibinfo{journal}{Phys. Rev. A} \textbf{\bibinfo{volume}{42}},
  \bibinfo{pages}{78} (\bibinfo{year}{1990}),
  \urlprefix\url{https://link.aps.org/doi/10.1103/PhysRevA.42.78}.

\bibitem[{\citenamefont{Adler}(2000)}]{Adler:2000aa}
\bibinfo{author}{\bibfnamefont{S.~L.} \bibnamefont{Adler}},
  \bibinfo{journal}{Physics Letters A} \textbf{\bibinfo{volume}{265}},
  \bibinfo{pages}{58} (\bibinfo{year}{2000}).

\bibitem[{\citenamefont{Bauer et~al.}(2013)\citenamefont{Bauer, Bernard, and
  Tilloy}}]{Bauer:2013aa}
\bibinfo{author}{\bibfnamefont{M.}~\bibnamefont{Bauer}},
  \bibinfo{author}{\bibfnamefont{D.}~\bibnamefont{Bernard}}, \bibnamefont{and}
  \bibinfo{author}{\bibfnamefont{A.}~\bibnamefont{Tilloy}},
  \bibinfo{journal}{Phys. Rev. A} \textbf{\bibinfo{volume}{88}},
  \bibinfo{pages}{062340} (\bibinfo{year}{2013}),
  \urlprefix\url{https://link.aps.org/doi/10.1103/PhysRevA.88.062340}.

\bibitem[{\citenamefont{Bauer et~al.}(2017)\citenamefont{Bauer, Bernard, and
  Jin}}]{Bauer:2017aa}
\bibinfo{author}{\bibfnamefont{M.}~\bibnamefont{Bauer}},
  \bibinfo{author}{\bibfnamefont{D.}~\bibnamefont{Bernard}}, \bibnamefont{and}
  \bibinfo{author}{\bibfnamefont{T.}~\bibnamefont{Jin}},
  \bibinfo{journal}{SciPost Physics} \textbf{\bibinfo{volume}{3}},
  \bibinfo{pages}{033} (\bibinfo{year}{2017}), \eprint{arXiv:1706.03984}.

\bibitem[{\citenamefont{Chenu et~al.}(2017)\citenamefont{Chenu, Beau, Cao, and
  Del~Campo}}]{Chenu:2017aa}
\bibinfo{author}{\bibfnamefont{A.}~\bibnamefont{Chenu}},
  \bibinfo{author}{\bibfnamefont{M.}~\bibnamefont{Beau}},
  \bibinfo{author}{\bibfnamefont{J.}~\bibnamefont{Cao}}, \bibnamefont{and}
  \bibinfo{author}{\bibfnamefont{A.}~\bibnamefont{Del~Campo}},
  \bibinfo{journal}{Physical Review Letters} \textbf{\bibinfo{volume}{118}},
  \bibinfo{pages}{140403} (\bibinfo{year}{2017}).

\bibitem[{\citenamefont{Bernhart}(1999)}]{Bernhart:1999aa}
\bibinfo{author}{\bibfnamefont{F.~R.} \bibnamefont{Bernhart}},
  \bibinfo{journal}{Discrete Mathematics} \textbf{\bibinfo{volume}{204}},
  \bibinfo{pages}{73 } (\bibinfo{year}{1999}).

\bibitem[{\citenamefont{Sloane}(2017)}]{Sloane:2017aa}
\bibinfo{author}{\bibfnamefont{N.~J.~A.} \bibnamefont{Sloane}},
  \emph{\bibinfo{title}{The on-line encyclopedia of integer sequences}}
  (\bibinfo{year}{2017}), \urlprefix\url{https://oeis.org}.

\bibitem[{\citenamefont{Andrews and Thirunamachandran}(1977)}]{Andrews:1977aa}
\bibinfo{author}{\bibfnamefont{D.~L.} \bibnamefont{Andrews}} \bibnamefont{and}
  \bibinfo{author}{\bibfnamefont{T.}~\bibnamefont{Thirunamachandran}},
  \bibinfo{journal}{The Journal of Chemical Physics}
  \textbf{\bibinfo{volume}{67}}, \bibinfo{pages}{5026} (\bibinfo{year}{1977}).

\bibitem[{\citenamefont{Links et~al.}(2003)\citenamefont{Links, Zhou, McKenzie,
  and Gould}}]{Links:2003}
\bibinfo{author}{\bibfnamefont{J.}~\bibnamefont{Links}},
  \bibinfo{author}{\bibfnamefont{H.-Q.} \bibnamefont{Zhou}},
  \bibinfo{author}{\bibfnamefont{R.~H.} \bibnamefont{McKenzie}},
  \bibnamefont{and} \bibinfo{author}{\bibfnamefont{M.~D.} \bibnamefont{Gould}},
  \bibinfo{journal}{Journal of Physics A: Mathematical and General}
  \textbf{\bibinfo{volume}{36}}, \bibinfo{pages}{R63} (\bibinfo{year}{2003}).

\bibitem[{\citenamefont{Gumbel}(2012)}]{Gumbel:2012aa}
\bibinfo{author}{\bibfnamefont{E.~J.} \bibnamefont{Gumbel}},
  \emph{\bibinfo{title}{Statistics of extremes}} (\bibinfo{publisher}{Courier},
  \bibinfo{address}{North Chelmsford, MA}, \bibinfo{year}{2012}).

\end{thebibliography}

\end{document}